\documentclass[twocolumn]{article}

\usepackage{arxiv}

\usepackage[utf8]{inputenc} 
\usepackage[T1]{fontenc}    
\usepackage{hyperref}       
\usepackage{url}            
\usepackage{booktabs}       
\usepackage{amsfonts}       
\usepackage{amsmath}
\usepackage{amssymb}
\usepackage{nicefrac}       
\usepackage{microtype}      
\usepackage{cleveref}       
\usepackage{graphicx}
\usepackage[square,numbers,sort]{natbib}
\usepackage{doi}
\usepackage{capt-of}
\usepackage{subcaption}
\usepackage{multirow}
\usepackage{abstract}
\usepackage{balance}

\title{Neural Proxies for Sound Synthesizers: Learning Perceptually Informed Preset Representations}

\date{}

\author{\href{https://orcid.org/0009-0002-6930-5980}{\includegraphics[scale=0.06]{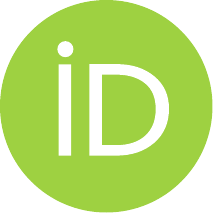}}\hspace{1mm}Paolo Combes$^{1}$\thanks{This is the Author’s Accepted Manuscript (AAM), accepted for publication in the Journal of the Audio Engineering Society.\\Author to whom correspondence should be addressed, email: \href{mailto:paolocombes@gmail.com}{paolocombes@gmail.com}} \quad \href{https://orcid.org/0000-0002-9427-9932}{\includegraphics[scale=0.06]{assets/orcid.pdf}}\hspace{1mm} Stefan Weinzierl$^1$ \quad Klaus Obermayer$^2$\\
        $^1$Audio Communication Group, TU Berlin, Berlin, Germany\\
        $^2$Neural Information Processing Group, TU Berlin, Berlin, Germany}

\renewcommand{\headeright}{Combes et al.}

\renewcommand{\shorttitle}{Neural Proxies for Sound Synthesizers}

\hypersetup{
pdftitle={Neural Proxies for Sound Synthesizers: Learning Perceptually Informed Preset Representations},
pdfsubject={cs.LG, cs.SD, eess.AS},
pdfauthor={Paolo Combes, Stefan Weinzierl, Klaus Obermayer},
pdfkeywords={First keyword, Second keyword, More},
}
\begin{document}
\twocolumn[ 
    \begin{@twocolumnfalse} 
    \maketitle
    \begin{abstract}
        Deep learning appears as an appealing solution for Automatic Synthesizer Programming (ASP), which aims to assist musicians and sound designers in programming sound synthesizers. However, integrating software synthesizers into training pipelines is challenging due to their potential non-differentiability. This work tackles this challenge by introducing a method to approximate arbitrary synthesizers. Specifically, we train a neural network to map synthesizer presets onto an audio embedding space derived from a pretrained model. This facilitates the definition of a neural proxy that produces compact yet effective representations, thereby enabling the integration of audio embedding loss into neural-based ASP systems for black-box synthesizers. We evaluate the representations derived by various pretrained audio models in the context of neural-based nASP and assess the effectiveness of several neural network architectures -- including feedforward, recurrent, and transformer-based models -- in defining neural proxies. We evaluate the proposed method using both synthetic and hand-crafted presets from three popular software synthesizers and assess its performance in a synthesizer sound matching downstream task. While the benefits of the learned representation are nuanced by resource requirements, encouraging results were obtained for all synthesizers, paving the way for future research into the application of synthesizer proxies for neural-based ASP systems.
    \end{abstract}
    \vspace{1em}
    \end{@twocolumnfalse} 
] 

\saythanks

\setcounter{section}{-1}  

\section{Introduction}
\label{sec:intro}
\subsection{Background}
Sound synthesizers have revolutionized music by expanding the timbral palette available to musicians, enabling the creation of entirely new sounds. However, the vast creative possibilities offered by synthesizers introduce significant usability challenges. Synthesizer interfaces are typically designed around their underlying synthesis methods, with control parameters directly mapping to the internal workings of the synthesis engine \cite{seagoAnalysisSynthesizerUser2004}. As a result, these interfaces necessitate non-intuitive parameter adjustments for achieving conceptually simple sound manipulations \cite{krekovicInsightsHabitsAttitudes2019}. As synthesis methods become more complex, synthesizer usability decreases, reducing accessibility for a wider range of musicians and sound designers.

To address these usability challenges, researchers have developed a range of systems for Automatic Synthesizer Programming (ASP), which refers to any methods aiming to support the process of programming a sound synthesizer \cite{shierSynthesizerProgrammingProblem2021}. These systems include synthesizer sound matching methods \cite{yee-kingAutomaticProgrammingVST2018,eslingFlowSynthesizerUniversal2019, chenSound2SynthInterpretingSound2022, vaillantImprovingSynthesizerProgramming2021}, query from vocal imitation \cite{mcartwright2014}, high-level control using timbral attributes \cite{krekovicAlgorithmControllingArbitrary2016} or natural language \cite{cherepCreativeTexttoAudioGeneration2024}. 
Other methods include techniques for interpolation \cite{levaillantLatentSpaceInterpolation2024}, generation \cite{peacheyCreatingLatentRepresentations2023}, and retrieval \cite{bradeSynthScribe2024} of presets -- pre-programmed lists of synthesizer parameter values used to generate sound.

Recent ASP systems typically rely on either genetic algorithms or deep learning approaches. While methods based on genetic algorithms can achieve high prediction quality, they are prohibitively time-consuming for interactive use in creative contexts \cite{masudaQualityDiversitySynthesizer2021}. In contrast, neural-based methods for Automatic Synthesizer Programming (nASP) offer an attractive solution, as they provide a balanced trade-off between prediction quality and computational efficiency. Additionally, their greater flexibility makes them more suitable for creative applications. 

\subsection{Challenges in nASP systems}
Two primary challenges arise when applying deep learning to synthesizers.

First, integrating synthesizers into deep learning training pipelines is inherently challenging. Neural networks are typically trained using gradient-based optimization techniques. Although effective, this approach limits the integration of non-differentiable synthesizers with unknown internal structure -- i.e., black-box synthesizers -- within an end-to-end training pipeline.
To maintain the differentiability required for gradient-based optimization, the synthesizer must be positioned after the loss computation within the training pipeline. As a result, the loss function can only depend on synthesizer parameter values, since incorporating the synthesizer into the network would disrupt the backpropagation chain.
This poses a challenge since loss functions based on synthesizer parameter values -- rather than synthesized audio -- are suboptimal for two reasons. First, parameter differences are not necessarily proportional to perceptual differences, which diminishes nASP system performance in tasks requiring perceptual evaluation. Second, in supervised training, this requires the ground truth to be based on synthesizer parameter values, thereby restricting training data to in-domain sounds, i.e., sounds produced by the synthesizer itself, raising concerns about the generalization ability of nASP systems to arbitrary sounds.

The second challenge arises from the data-driven approach required to train nASP systems. Ideally, in-domain training datasets should be derived from hand-crafted presets, i.e., presets designed by humans, which is both costly and time-consuming. As a result, the number of available hand-crafted presets is usually scarce. This scarcity hinders the use of larger and more powerful architectures, since they typically require more training data. Moreover, this makes the use of a specific synthesizer dependent on the available number of hand-crafted presets, which is a considerable drawback.
Therefore, these two key challenges limit the performance of nASP methods across arbitrary synthesizers and must be addressed.

\subsection{Main Contributions}
In this work, we propose a method to tackle the challenge posed by the lack of differentiability using neural proxies, which are neural networks trained to approximate black-box functions \cite{jacoviNeuralNetworkGradientbased2019}. Neural proxies have been used to model various audio effects, including guitar amplifiers and effects chains \cite{wrightRealTimeGuitarAmplifier2020, steinmetzStyleTransferAudio2022}. Our approach aims to train a half-hybrid neural proxy \cite{steinmetzStyleTransferAudio2022}, where the proxy is used during training, but the software synthesizer is used during inference. More specifically, we propose the following:
\begin{itemize}
	\item A method for training a neural proxy for arbitrary synthesizers, referred to as the preset encoder. The preset encoder is trained to map synthesizer presets onto an audio embedding space, defined by a pretrained audio model serving as a fixed feature extractor. 
	\item A sound attribute ranking evaluation to assess representations derived from pretrained audio models for defining loss functions in nASP systems.
    \item A benchmark comparing the impact of various neural network architectures and input preprocessing techniques on the preset encoder's performance across three synthesizers of varying complexity.
    \item An analysis of the generalization ability of preset encoders trained on a large-scale dataset of 10M synthetic presets to hand-crafted presets, examining how synthesizer programming complexity affects generalization.
	\item An evaluation of the proposed method on a synthesizer sound matching downstream task, demonstrating its integration into nASP systems and highlighting its benefits.
\end{itemize}

Once trained, the preset encoder enables the use of audio embedding loss functions -- loss functions based on representations derived from pretrained audio models -- for training nASP systems. By doing so, we aim to facilitate the development of neural proxies for training nASP systems, thereby improving the performance of deep learning models for arbitrary synthesizers. This, in turn, improves synthesizer usability and enables a broader range of music creators and sound designers to translate their ideas into sound more effectively.

While prior work \cite{barkanInversynthIISound2023} has demonstrated the value of auditory loss functions derived from hand-crafted time-frequency representations produced by a neural proxy, we propose a complementary approach by integrating losses derived from pretrained audio models. Our initial findings align with previous research \cite{barkanInversynthIISound2023, masudaImprovingSemiSupervisedDifferentiable2023, uzradDiffMoogDifferentiableModular2024}, supporting the notion that incorporating synthesized audio-based losses provides advantages over using parameter value-based losses alone. In contrast to the aforementioned studies, which use large, hand-crafted representations, our results suggest that improved performance in sound matching tasks can also be achieved using more compact representations that integrate both hand-crafted and learned features -- offering a promising direction for efficient representations in nASP systems that warrants further exploration.

However, while our experiments demonstrate the potential benefits of representations derived from pretrained audio models, we acknowledge certain limitations regarding their practicality and task suitability. A more detailed discussion of these limitations can be found in Sec.~\ref{sec:limitations}. Furthermore, our analysis of the generalization ability on hand-crafted presets highlights the limitations of training solely on synthetic presets, even at a large scale, and suggests that further research in this area would provide valuable insights.

The source code and data required to reproduce the results presented in this paper are available in the accompanying repository.\footnote{\url{https://github.com/pcmbs/synth-proxy}}

The rest of this paper is organized as follows. Sec.~\ref{sec:related} presents the related work. Sec.~\ref{sec:method} details the proposed method. Sec.~\ref{sec:shared_design} describes the experimental design shared across all experiments. Sec.~\ref{sec:am_select} details the sound attribute ranking evaluation for the pretrained audio model. Sec.~\ref{sec:sp_eval} assesses the preset encoders' performance on synthetic presets, while Sec.~\ref{sec:hc_eval} examines performance on hand-crafted presets and discusses generalization ability. Sec.~\ref{sec:ssm_eval} evaluates the proposed method on a synthesizer sound matching downstream task. Finally, Sec.~\ref{sec:limitations} discusses limitations of learned representations, and Sec.~\ref{sec:conclusion} summarizes our findings.

\section{Related Work}
\label{sec:related}

The motivation for using neural proxies stems from the recent development of differentiable synthesizers using automatic differentiation frameworks, which have enabled the use of auditory loss functions based on synthesized audio \cite{engelDDSPDifferentiableDigital2020}.
Incorporating auditory loss in nASP systems for synthesizer sound matching has yielded promising results compared to relying solely on parameter loss \cite{barkanInversynthIISound2023, masudaImprovingSemiSupervisedDifferentiable2023, uzradDiffMoogDifferentiableModular2024} and has further enabled the training of nASP systems on out-of-domain sounds \cite{caspeDDX7DifferentiableFM2022,Yang:2023:WBS}.

Numerical gradient approximation methods have been investigated as alternatives to neural proxies to derive auditory loss functions. A method based on finite differences for nASP has been proposed in \cite{chenSound2SynthInterpretingSound2022}. However, it required rendering audio twice for each synthesizer parameter of every training preset and was omitted from the final model due to its high computational cost. 
A more efficient technique, Simultaneous Perturbation Stochastic Approximation (SPSA) \cite{spallOverviewSimultaneousPerturbation1998}, which reduces the number of required audio renderings per iteration to two, has been used for automatic audio effects programming \cite{martinezramirezDifferentiableSignalProcessing2021}. This approach was compared with neural proxies for audio effects in \cite{steinmetzStyleTransferAudio2022}, which found that SPSA was more susceptible to training instability and required careful hyperparameter tuning. 
Furthermore, both studies only evaluated the method on continuous numerical parameters, and it is unclear how to address perturbations of categorical and binary parameters, which might cause further instability.

In closely related work, Inversynth2 \cite{barkanInversynthIISound2023} is among the first approaches to employ a neural proxy for black-box synthesizers within an nASP system for synthesizer sound matching.
The neural proxy replaces the true synthesizer during training and produces time-frequency hand-crafted features, enabling the computation of an auditory loss function.
The key difference in our approach lies in the introduction of a loss function derived from more compact representations derived from pretrained audio models, rather than solely relying on hand-crafted features. This further explores alternatives to using a parameter loss alone.
SPINVAE and its successor, SPINVAE-2, from another line of research, utilized a transformer-based VAE model for preset interpolation on the Dexed synthesizer \cite{vaillantSynthesizerPresetInterpolation2023, levaillantLatentSpaceInterpolation2024}. While SPINVAE required synthesizing audio to produce latent representations, SPINVAE-2 avoided this non-differentiable operation, allowing the gradients to be backpropagated from the latent representation to the encoder's input, thus enabling the formulation of a differentiable synthesizer proxy. 
A distinguishing feature of our approach is its flexibility to incorporate advancements in audio representation learning, potentially leveraging more diverse, compact, and efficient representations. 
A comparison of the proposed method with Inversynth2 and SPINVAE-2 for producing the input representation of a loss function is left for future work.

\section{Methodology}
\label{sec:method}
\subsection{Preset Formalization}
\label{ssec:preset_form}
We define a preset as a vector of parameters $\underline{x} \in \mathcal{P}_s$, where $\mathcal{P}_s$ denotes the parameter space of a given synthesizer $s$ with a total number of MIDI controllable parameters $N_s:=\operatorname{dim}\left(\mathcal{P}_s\right)$. In most cases, synthesizers are characterized by a combination of three types of parameters: numerical parameters $x_i \! \in \! [0,1], \forall i \! \in \! I_{\text{num}}^s$; binary parameters $x_j \! \in \! \{0,1\}, \forall j \! \in \! I_{\text{bin}}^s$; and categorical parameters whose categories are represented by an integer $x_k \! \in \! \{1, \ldots, \left|\mathcal{C}_k^s\right|\}, \forall k \! \in \! I_{\text{cat}}^s$, where $\left|\mathcal{C}_k^s\right|$ denotes the number of categories of parameter $k$. Here, $I_{\text{num}}^s$, $I_{\text{bin}}^s$ and $I_{\text{cat}}^s$ denote the index sets of numerical, binary, and categorical parameters, respectively, such that $\vert I_{\text{num}}^s \cup I_{\text{bin}}^s \cup I_{\text{cat}}^s \vert = N_s$. Hence, we can formally define $\mathcal{P}_s$ as:
\begin{multline}
  \mathcal{P}_s := \biggl\{ \underline{x} \in \mathbb{R}^{N_s} \mid x_i \in \left[0,1\right], \forall i \in I_{\text{num}}^s \\
  \wedge x_j \in \{0,1\}, \forall j \in I_{\text{bin}}^s \wedge x_k \in \{c\}_{c=1}^{\left| \mathcal{C}_k^s \right|}, \forall k \! \in \! I_{\text{cat}}^s \biggr\}
\end{multline}

Furthermore, we define a synthesizer as a function $s: \mathcal{P}_s \times \Omega \rightarrow \mathbb{R}^{n}, s(\underline{x}, \underline{\omega})=\underline{x}_a$ where $\underline{x}_a \! \in \! \mathbb{R}^{n}$ is an output audio vector of $n$ samples for a preset $\underline{x} \! \in \! \mathcal{P}_s$ and MIDI parameters $\underline{\omega} \! \in \! \Omega$ used to render the audio, i.e., the MIDI note, MIDI velocity, MIDI length, and duration in seconds.

\subsection{Proposed Method}
\label{ssec:method}

\begin{figure}[t]
  \centerline{\includegraphics[width=\linewidth]{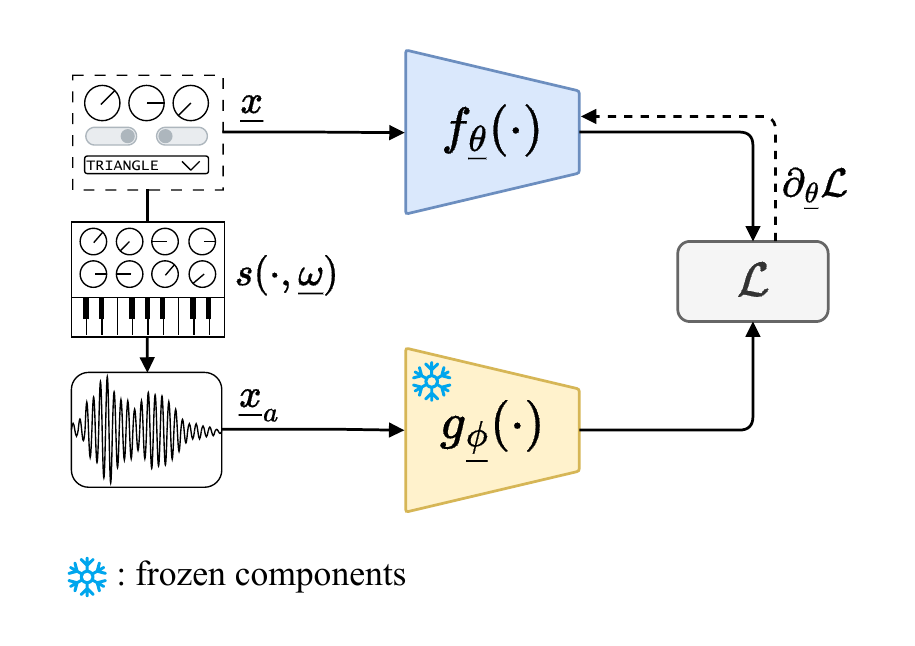}}
  \caption{\label{fig:model}{Overview of the proposed method. Given a preset $\underline{x} \in \mathcal{P}_s$ from a synthesizer $s$, a preset encoder $f_{\underline{\theta}}$ learns to minimize the distance between its representation of $\underline{x}$ and that produced by a pretrained audio model $g_{\underline{\phi}}$ derived from the synthesized audio $s(\underline{x}, \underline{\omega})=\underline{x}_a$ with MIDI parameters $\underline{\omega} \in \Omega$.}}
\end{figure}

This work introduces a method for training a neural proxy for an arbitrary black-box synthesizer, as depicted in Fig.~\ref{fig:model}. This method relies on two neural networks: a preset encoder and a pretrained audio model. The preset encoder $f_{\underline{\theta}}: \mathcal{P}_s \rightarrow \mathbb{R}^{m}$, parametrized by $\underline{\theta} \in \Theta$, takes as input a preset $\underline{x} \in \mathcal{P}_s$ from a synthesizer $s$, while the pretrained audio model $g_{\underline{\phi}}: \mathbb{R}^{n} \rightarrow \mathbb{R}^{m}$, parametrized by $\underline{\phi} \in \Phi$, maps the synthesized audio $\underline{x}_a \in \mathbb{R}^{n}$ onto an $m$-dimensional audio embedding space.

The objective is to train the preset encoder to align its representations with those learned by the pretrained audio model, which is used as a fixed feature extractor. This approach aims to leverage the learned audio representations in a cross-modal knowledge distillation process, generating similar representations for presets that produce similar sounds according to the underlying audio model. Thus, the preset encoder is trained to act as a neural proxy for a given synthesizer. The optimization problem for a synthesizer $s$ can be formulated as follows:
\begin{equation}
  \min _{\underline{\theta} \in \Theta} d\left(f_{\underline{\theta}}(\underline{x}), g_{\underline{\phi}}\left(\underline{x}_a\right)\right) \triangleq \min _{\underline{\theta} \in \Theta} d\left(f_{\underline{\theta}}(\underline{x}), g_{\underline{\phi}}(s(\underline{x}, \underline{\omega}))\right)
\end{equation}
for an input preset $\underline{x} \in \mathcal{P}_s$, MIDI parameters $\underline{\omega} \in \Omega$, and where $d:=L^1$ was chosen as the distance function to compute the loss between the preset and audio representations, as it has been shown to outperform the $L^2$ distance in its ability to discriminate between audio representations \cite{turianOneBillionAudio2021}. The training objective for the optimization problem is thus defined as:
\begin{equation}
  \mathcal{L}(\underline{\theta}):=\mathbb{E}_{\underline{x}}\left[\left\|f_{\underline{\theta}}(\underline{x})-g_{\underline{\phi}}(s(\underline{x}, \underline{\omega}))\right\|_1\right].
\end{equation}

\section{Experimental Design}
\label{sec:shared_design}

\subsection{Datasets}
\label{ssec:datasets}

\subsubsection{Synthesizers}
\label{ssec:synths}
The evaluation of the preset encoders was conducted using three popular software synthesizers, each with distinct synthesis types and programming complexity levels.

\emph{Dexed}.\footnote{\url{https://github.com/asb2m10/dexed}} An open-source Frequency Modulation (FM) synthesizer that models the Yamaha DX7. While difficult to manipulate, it can produce a wide range of sounds with a relatively small number of parameters -- a characteristic property of FM synthesis \cite{mirandaComputerSoundDesign2002}. Among these, the algorithm parameter plays a particularly influential role, as it defines the routing of the six FM operators and can significantly affect both the synthesizer’s behavior and the relevance of other parameters.

\emph{Diva}. A Virtual Analog (VA) synthesizer developed by u-he,\footnote{\url{https://u-he.com/products/diva/}} characterized by a higher number of parameters, some of which have meanings that are not unique and can vary depending on the selected mode. In fact, the functionality of some parameters may vary or become irrelevant depending on the mode selected for each module. This interdependency highly affects the synthesized sound, particularly in the oscillator module.

\emph{TAL-NoiseMaker (TAL-NM)}. A simpler VA synthesizer developed by TAL-Software.\footnote{\url{https://tal-software.com/products/tal-noisemaker}} It is used as a baseline due to its fewer parameters and straightforward design, as well as to generate the dataset used to evaluate the pretrained audio models in Sec.~\ref{sec:am_select}.

\subsubsection{Synthetic Preset Datasets}
\label{sssec:data_synthetic}

\begin{table}[t]
  \caption{Number of parameters used per synthesizer, grouped by parameter type.}
  \centering
  \begin{tabular}{@{\extracolsep{4pt}}lrrrr@{}}
    \hline
    Synthesizer & \# Num. & \# Bin. & \# Cat. & Total \\
    \hline
    Dexed       & 92      & 0       & 8       & 100   \\
    Diva        & 100     & 22      & 51      & 173   \\
    TAL-NM      & 48      & 3       & 8       & 59    \\
    \hline
  \end{tabular}
  \label{tab:synths}
\end{table}
The training, validation, and test datasets were constructed by randomly sampling presets for each synthesizer using different random seeds. Continuous numerical parameters were sampled from a uniform distribution between 0 and 1, while discrete parameters -- categorical, binary, or discretized numerical -- were sampled from a categorical distribution.

Several synthesizer parameters were excluded and set to fixed values during rendering. This included the output volume, any parameters dependent on user interactions with a MIDI controller, polyphonic or MIDI note-related parameters, and parameters generating non-deterministic control rate behavior, such as the Sample \& Hold (S\&H) waveform for Low Frequency Oscillators (LFOs).

Depending on the specific synthesizer, parameters or parameter values deemed non-essential for the study were also omitted. The resulting total number of used parameters for each synthesizer is presented in Table \ref{tab:synths}. Further details on excluded parameters are available in the accompanying source code repository.

All audio was rendered using DawDreamer \cite{braunDawdreamerBridgingGap2021} for a duration of 5 seconds and converted to mono using a MIDI note of 60 (C3), played at a velocity of 100 for a duration of two seconds. Only presets producing audio within a specified RMS range were included in the datasets to mitigate the risk of generating excessively quiet or loud presets, which could lead to NaN outputs from the pretrained audio model. To accelerate the training process, the audio representations produced by the pretrained audio model were precomputed.

While the use of a single MIDI note and velocity may pose a limitation, this approach was deemed sufficient for the purposes of this proof-of-concept, as most ASP methods focused on black-box synthesizers are trained with a fixed MIDI note and velocity \cite{yee-kingAutomaticProgrammingVST2018,eslingFlowSynthesizerUniversal2019,chenSound2SynthInterpretingSound2022,barkanInversynthIISound2023,brufordSynthesizerSoundMatching2024,romaSoundMatchingUsing2024,levaillantLatentSpaceInterpolation2024}. Nonetheless, the current method can be easily extended to support multiple MIDI notes and velocities by treating them as additional synthesizer parameters, which can then be input alongside the other parameters into the preset encoder.

\subsubsection{Hand-Crafted Preset Datasets}
\label{sssec:hc_dataset}

\begin{table*}[!ht]
  \caption{Overview of the evaluated preset encoders. $\operatorname{PE_{sin}}$ denotes the sinusoidal positional encoding and $d_\text{hidden}$ denotes the encoder's hidden dimension. The number of layers reported excludes the projection head. The number of parameters is reported for Dexed, Diva, and TAL-NM, respectively. The variation in the number of parameters for a given preset encoder is attributed to the different number of synthesizer parameters used, which influences the dimensionality of the input layer and/or the number of parameters in the Preset Tokenizer.}
  \centering
  \begin{tabular}{@{\extracolsep{2pt}}lrrrrrrr@{}}
    \hline
    Encoder  & Input representation                                                                                           & $d_\text{token}$ & \# Layers & $d_{\text{hidden}}$ & \# Params (M)   \\
    \hline
    MLP-OH   & $\operatorname{OH}(\underline{x})$                                                                             & -                & 2         & 2048                & [4.9, 5.4, 4.8] \\
    HN-OH    & $\operatorname{OH}(\underline{x})$                                                                             & -                & 6         & 768                 & [6.2, 6.4, 6.1] \\
    HN-PT    & $\operatorname{Flatten}(\operatorname{PT}(\underline{x}))$                                                     & 64               & 6         & 512                 & [6.0, 8.4, 4.7] \\
    HN-PTGRU & $\operatorname{BiGRU}(\operatorname{PE_{sin}}(\operatorname{PT}(\underline{x})))$                              & 384              & 6         & 768                 & [7.1, 7.2, 7.1] \\
    TFM      & $\operatorname{PE_{sin}}(\operatorname{Concat}(\underline{w}_{\text{CLS}}, \operatorname{PT}(\underline{x})))$ & 256              & 6         & 256                 & [4.8, 4.9, 4.8] \\
    \hline
  \end{tabular}
  \label{tab:arch}
\end{table*}

The datasets of hand-crafted presets for Dexed, Diva, and TAL-NM were obtained from \cite{vaillantImprovingSynthesizerProgramming2021}, \cite{eslingFlowSynthesizerUniversal2019}, and online sources, respectively. Only the synthesizer parameters used during training were considered; the remaining parameters were set to their default values and excluded. After removing silent and duplicate presets, the final datasets comprised 27,609 presets for Dexed, 9,000 for Diva, and 404 for TAL-NM. 

Due to its limited size, TAL-NM was excluded from both the preset encoder evaluation on hand-crafted presets and the SSM downstream task evaluation.

\subsection{Preset Encoders}
\subsubsection{Preset Input Representation}
\label{ssec:preset_repr}

Depending on the backbone architecture, two different approaches were used to preprocess synthesizer presets.

\emph{One-Hot Encoding (OH)}. Numerical and binary parameters are represented using their raw values, in $\left[0,1\right]$ and in $\{0,1\}$, respectively, while categorical parameters are one-hot encoded. We use the term "one-hot encoding" to emphasize the distinction between this approach and using raw, unprocessed parameter values, although only a subset of the parameters (the categorical ones) are one-hot encoded.

\emph{Preset Tokenizer (PT)}. Each numerical and binary parameter is associated with a learned embedding vector $\underline{v}_i \in \mathbb{R}^{d_{\text{token}}}, \forall i \in I_{\text{num}}^s \cup I_{\text{bin}}^s$ with token dimension $d_{\text{token}}$. This embedding vector is learned during training as part of the model, meaning that the model adjusts these vectors to capture useful information about the parameters. The final representation, or "token," of a numerical parameter is the learned vector scaled by the actual parameter value. In the case of binary parameters, the token is either a zero vector or the learned embedding vector itself, depending on the binary value. Additionally, each categorical parameter is mapped to a lookup table $W_i \in \mathbb{R}^{\left|\mathcal{C}_i^s\right| \times d_\text{token}}, \forall i \in I_{\text{cat}}^s$. This lookup table contains embeddings for each category of the categorical parameter, which are also learned during training.  The preset tokenizer can thus be defined as:
\begin{equation}
  \operatorname{PT}: \mathcal{P}_s \rightarrow \mathbb{R}^{N_s \times d_\text{token}}, \operatorname{PT}(\underline{x}):=\left[\underline{t}_i(x_i) \right]_{i=1}^{N_s}
\end{equation}
with
\begin{equation}
  \underline{t}_i(x_i):=\left\{\begin{array}{ll}x_i \underline{v}_i & i \in I_{\text{num}}^s \cup I_{\text{bin}}^s \\ \underline{e}_{x_i}^{\top} W_i & i \in I_{\text{cat}}^s\end{array}\right. \in \mathbb{R}^{d_{\text{token}}},
\end{equation}
where $\underline{e}_{x_i} \in \mathbb{R}^{d_{\text{token}}}$ is the $x_i$-th standard unit vector used to retrieve the corresponding embedding for the categorical parameter $i$ from $W_i$.

This method corresponds to the one used in SPINVAE \cite{vaillantSynthesizerPresetInterpolation2023} and in several tabular data models such as FT-Transformer \cite{gorishniyRevisitingDeepLearning2021}. The presented method only differs in that binary parameters are treated as numerical rather than categorical, which reduces the number of weights required for binary parameters by half.

\subsubsection{Backbone Architectures}
\label{ssec:backbone}

Various combinations of input representations and backbone architectures were investigated for the preset encoders. An overview of the different preset encoders is presented in Table \ref{tab:arch}.

The baseline encoder, denoted MLP-OH, consists of a wide but shallow multilayer perceptron with a single hidden layer taking one-hot encoded presets as input. The output from the hidden layer is then projected onto the audio embedding space using a fully connected layer.
To enhance the representation capacity of the baseline, a Highway Network-based encoder (HN) \cite{srivastavaTrainingVeryDeep2015a} was implemented by increasing the number of layers, reducing the hidden dimension, and adding gated residual connections to aid optimization, resulting in the HN-OH encoder.

Further development resulted in the HN-PT encoder, which substitutes the one-hot encoding with a flattened representation of the tokenized synthesizer parameters as described in Sec.~\ref{ssec:preset_repr}. This change is intended to capture the meaning of each synthesizer parameter more effectively through learnable vectors.
To address the high input layer dimension $d_{\text{in}} \! = \! N_s \cdot d_{\text{token}}$ resulting from the flattening operation, the HN-PTGRU encoder was conceived. It compresses the sequence of tokens into a more compact representation using a Bidirectional Gated Recurrent Unit (BiGRU) layer with sinusoidal positional encoding, making the input representation independent of the number of synthesizer parameters. The final hidden states of both directions are concatenated and fed into a Highway Network.

Lastly, a transformer-based encoder, denoted TFM, was developed to leverage the attention mechanism's potential to better capture the interdependencies between synthesizer parameters. After the input preset tokenization, a special learnable token $\underline{w}_{\text{CLS}} \! \in \! \mathbb{R}^{d_{\text{token}}}$, intended for use as a final representation, is prepended to the sequence of tokens. Sinusoidal positional encoding is added before the sequence is fed into the encoder. The special token of the last layer undergoes layer normalization and a ReLU non-linearity before being projected onto the audio embedding space using a fully connected layer.

All encoders employed ReLU activations and batch normalization, except TFM, which utilized layer normalization. The architecture hyperparameters were chosen through hyperparameter optimization (HPO), as described in Appendix \ref{app:sp_hpo}, to ensure all models had a similar number of parameters, as shown in Table \ref{tab:arch}. The selected architecture hyperparameters for TFM mirrored those used in SPINVAE \cite{vaillantSynthesizerPresetInterpolation2023}, except with 8 attention heads instead of 4, which yielded better performance during HPO.

\section{Audio Model Ranking Evaluation}
\label{sec:am_select}
\subsection{Experimental Design}
\label{ssec:am_design}
Various pretrained audio models were evaluated to determine their suitability as reference models for the preset encoder. Specifically, we designed a ranking experiment to evaluate the ability of each model to order sounds based on monotonic changes in parameter values linked to different sound attributes. The goal was to evaluate the perceptual relevance of the audio embedding space by examining whether distances within the audio embedding space correlate with changes in sound attributes linked to human auditory perception. To this end, we created a dataset based on TAL-NM, which consists of 13 groups, each relating to a given synthesizer parameter. Each group contains 10 presets, and for each preset, the parameter associated with that group was monotonically increased in 20 steps to modify a specific sound attribute. These include parameters related to the perceived sound envelope (e.g., amp attack, amp decay), brightness (e.g., filter cutoff, filter resonance), and pitch (e.g., pitch coarse), which directly correspond to aspects of human auditory perception. For bipolar parameters -- parameters that can take on both positive and negative values, usually centered around zero -- we restricted all 20 values within each preset to be either greater or less than their midpoints.

For each sound attribute, we compute a ranking score as follows: (i) obtain the representation of each sound using the audio model under evaluation and apply a temporal reduction function across time frames; (ii) compute the distance matrix for each preset based on pairwise $L^1$ distances; (iii) for each preset, sort the sounds in ascending order based on their distances to the minimum and maximum parameter values, compute the Spearman rank correlation coefficients for both rankings, and take the average of these two correlations.

We evaluated seven popular model families, encompassing AudioMAE \cite{NEURIPS2022_b89d5e20}, CLAP \cite{wuLargescaleContrastiveLanguageAudio2024}, DAC \cite{NEURIPS2023_58d0e78c}, EfficientAT \cite{schmidLowComplexityAudioEmbedding2023} (denoted with the prefix \texttt{mn} to indicate the use of MobileNetV3 as a backbone), M2L \cite{pasiniMusic2LatentConsistencyAutoencoders2024}, OpenL3 \cite{cramerLookListenLearn2019}, and PaSST \cite{koutiniEfficientTrainingAudio2022}. Among these models, both the EfficientAT and PaSST model families use final representations that concatenate hand-crafted features (time-averaged Mel-spectrograms) with learned features, whereas the other model families rely solely on learned features in their final representations. A detailed description of each model family is provided in Appendix \ref{app:am_descr}. We also included several hand-crafted audio representations as baselines. For models producing a representation for each input timeframe rather than for the whole sequence, we evaluated two temporal reduction methods: (i) \texttt{nop}, in which no reduction is applied, and the final representation is the concatenation of outputs from each timeframe, and (ii) \texttt{avg time}, in which the mean of each timeframe’s representation is taken as the final representation, yielding a representation that is independent of the input audio length. For each model, we chose the reduction function yielding the overall best result. This procedure was applied to all models except CLAP, whose final representation is already time-averaged and is therefore denoted accordingly.

\subsection{Results}
\label{am_results}

\begin{figure*}[!ht]
  \centerline{\includegraphics[width=\textwidth]{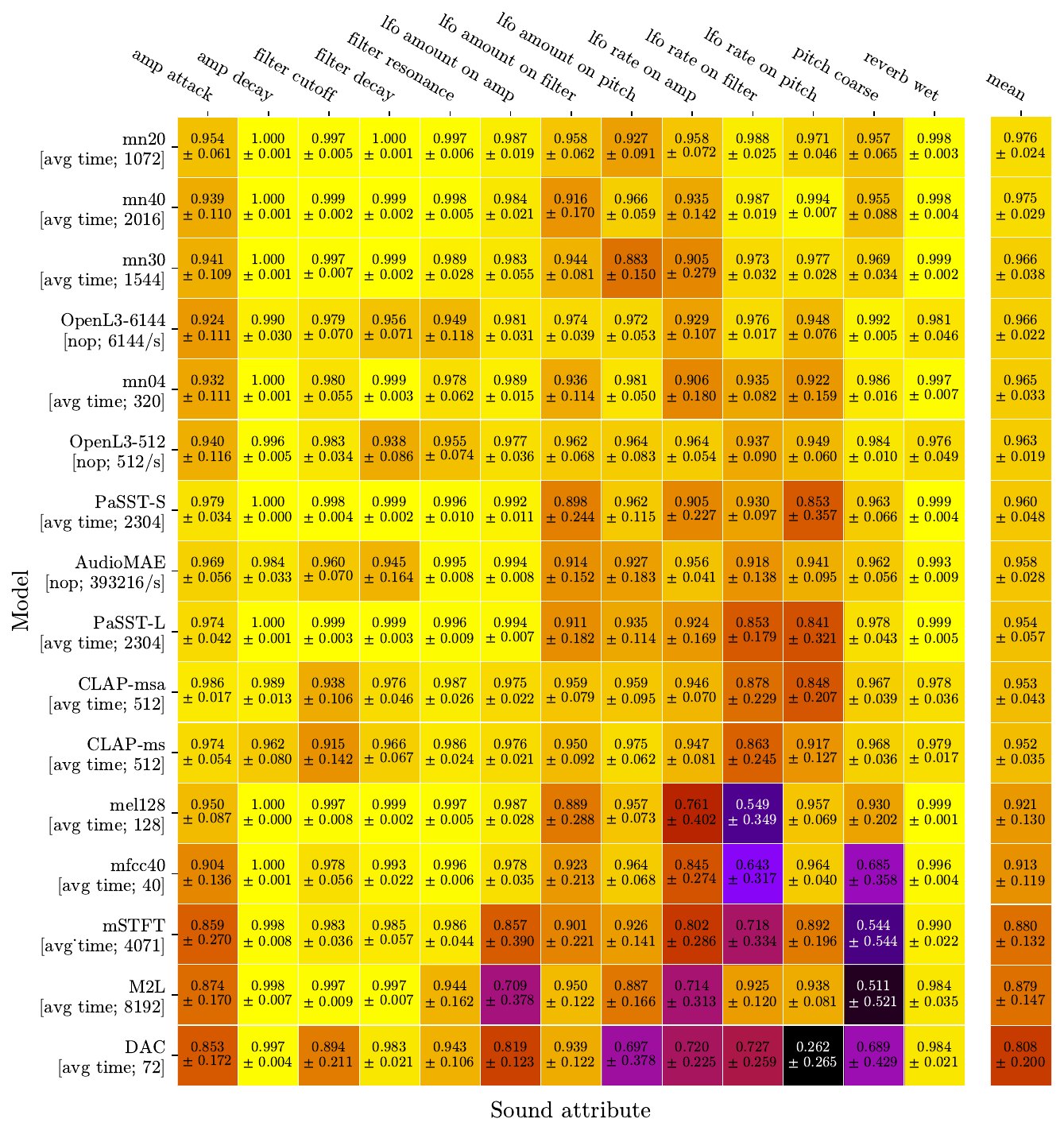}}
  \caption{Performance of each pretrained audio model on the ranking evaluation. The mean and standard deviation of the Spearman correlation, over the 20 rankings per synthesizer parameter, are reported. For each model, the temporal reduction function yielding the overall best result, as well as the final representation size, is indicated in bracket – which is indicated per second if no reduction, i.e., \texttt{nop} is used.}
  \label{fig:am_results}
\end{figure*}
The results of the ranking evaluation are summarized in Fig.\ref{fig:am_results}.

The EfficientAT model family demonstrated the best overall results, highlighting its robust performance. 
These findings, along with the results from the PaSST models, suggest that combining hand-crafted features with those extracted from models is an efficient approach that calls for further exploration.
Interestingly, time-averaged representations outperformed frame-flattened ones in most models. This suggests that retaining all features over time may not improve sound attribute ranking and that the temporal mean provides a summary statistic that might be more beneficial in most cases. This result is surprising, as sounds often evolve in complex ways when, e.g., temporal modulations or arpeggiators are employed.
It also appears that representation size, although impacting final performance, is not a crucial factor, as both strong and weak performances were observed across models with varying representation sizes.

Models trained on AudioSet -- including AudioMAE, EfficientAT, OpenL3, and PaSST -- achieved the highest scores, despite the dataset's lack of high-quality audio samples and the fact that only 50\% of its audio events are musical, which was unexpected.
Among these, OpenL3 is the best-performing model whose representation does not include hand-crafted features in its representation. In contrast, hand-crafted representations, used as baselines, generally underperformed compared to models leveraging learned features, aligning with the findings in \cite{turianOneBillionAudio2021}. 
Interestingly, the DAC and M2L autoencoders trained for audio compression yielded poor results, which may stem from a mismatch between their training objectives and the perceptual ranking task at hand.

At first glance, sampling rates do not appear to have a significant effect. For example, AudioMAE, trained with inputs sampled at 16 kHz, still achieved good results. However, it is important to note that most of these models were trained on AudioSet, which may not contain informative high-frequency content, preventing the models from fully exploiting this part of the spectrum.

Synthesizer parameters that influenced both temporal and frequency dimensions, such as LFO parameters, produced the highest variance in model results. In contrast, parameters affecting purely temporal (e.g., amplitude parameters) or purely frequential aspects (e.g., filter parameters) showed less variance. Surprisingly, all models performed well on the reverb wet parameter.
Notably, hand-crafted features struggled with pitch-coarse parameters, where learned representations significantly outperformed them.
Finally, lower-scoring models exhibited higher variances, indicating performance instability for certain synthesizer parameters.

Given the generally high scores achieved by the models, future work could focus on designing more challenging evaluation procedures, such as integrating additional synthesizers and varying multiple parameters simultaneously, to better assess model performance in complex scenarios. Additionally, examples targeting the higher end of the frequency spectrum could be used to better evaluate the significance of input sampling rates.

As mentioned previously, the best-performing models that exclude hand-crafted features tend to rely on large embeddings and achieve optimal results without temporal reduction -- conditions that significantly compromise their practicality. This raises concerns about the practicality of relying exclusively on learned representations and motivates further exploration of hybrid approaches for deriving compact and effective representations.

\section{Synthetic Presets Evaluation}
\label{sec:sp_eval}
\subsection{Experimental Design}
\label{ssec:sp_design}

\subsubsection{Datasets}
\label{sssec:sp_datasets}
The synthetic preset datasets were generated following the procedure described in Sec.~\ref{sssec:data_synthetic}. Each dataset example consists of a preset serving as the input to the preset encoder and the corresponding audio representation as the target. The dataset sizes were set to 10M examples for training, 131K for validation, and 131K for testing. For HPO, separate training and validation datasets of 1M and 64K examples were generated. The size of the training dataset was selected to assess whether training on a sufficiently large number of synthetic presets could mitigate the data scarcity issue by enabling the preset encoder to generalize to hand-crafted presets.

Given the large scale of the training dataset, we prioritized practicality over performance by selecting a higher-throughput model that produces more compact representations while still achieving acceptable results. The computational costs and storage requirements of better models were deemed too high for generating the datasets for the three synthesizers. Specifically, we selected the \texttt{mn04} variant of the EfficientAT models with the \texttt{avg time} reduction function, which offers a good trade-off between representation size and performance. To further reduce the representation size, we discarded the hand-crafted Mel features, resulting in a 192-dimensional representation independent of the input audio length. This model achieved an overall mean score of 0.928 $\pm$ 0.055 in the ranking evaluation, remaining competitive despite reduced performance, and is further discussed in Sec.~\ref{sec:limitations}. As a result, we reduced the dataset size to approximately 8 GB for each synthesizer, compared to 44 GB if the best-performing model -- \texttt{mn20} with the \texttt{avg time} reduction -- had been used, which was deemed too high. 

\begin{table*}[!ht]
  \caption{Comparison between models on synthetic presets. The mean and standard deviation of the MRR and average $L^1$ error, over 100 runs, are reported. Both metrics are calculated on 4096 randomly chosen synthetic presets per run. The best-performing models for both metrics are highlighted in bold.}
  \centering
  \begin{tabular}{@{\extracolsep{4pt}}lcccccc@{}}
    \hline
    \multicolumn{1}{c}{} & \multicolumn{2}{c}{Dexed}             & \multicolumn{2}{c}{Diva}               & \multicolumn{2}{c}{TAL-NM}                                                                                                                                      \\
    \cline{2-3} \cline{4-5} \cline{6-7}
    Model                & MRR ($\uparrow$)                      & $L^1$ Error ($\downarrow$)             & MRR ($\uparrow$)                      & $L^1$ Error ($\downarrow$)             & MRR ($\uparrow$)                      & $L^1$ Error ($\downarrow$)             \\
    \hline
    MLP-OH               & 0.229 \scriptsize{$\pm$6e-3}          & 0.0953 \scriptsize{$\pm$7e-4}          & 0.216 \scriptsize{$\pm$5e-3}          & 0.0734 \scriptsize{$\pm$5e-4}          & 0.677 \scriptsize{$\pm$6e-3}          & 0.0719 \scriptsize{$\pm$6e-4}          \\
    HN-OH                & 0.665 \scriptsize{$\pm$7e-3}          & 0.0605 \scriptsize{$\pm$6e-4}          & 0.605 \scriptsize{$\pm$7e-3}          & 0.0547 \scriptsize{$\pm$5e-4}          & 0.937 \scriptsize{$\pm$3e-3}          & 0.0434 \scriptsize{$\pm$4e-4}          \\
    HN-PT                & 0.585 \scriptsize{$\pm$7e-3}          & 0.0665 \scriptsize{$\pm$6e-4}          & 0.561 \scriptsize{$\pm$7e-3}          & 0.0645 \scriptsize{$\pm$6e-3}          & 0.923 \scriptsize{$\pm$3e-3}          & 0.0566 \scriptsize{$\pm$1e-2}          \\
    HN-PTGRU             & 0.794 \scriptsize{$\pm$6e-3}          & 0.0507 \scriptsize{$\pm$8e-4}          & 0.628 \scriptsize{$\pm$6e-3}          & 0.0531 \scriptsize{$\pm$4e-4}          & 0.956 \scriptsize{$\pm$3e-3}          & 0.0442 \scriptsize{$\pm$5e-3}          \\
    TFM                  & \textbf{0.872} \scriptsize{$\pm$5e-3} & 0\textbf{.0435} \scriptsize{$\pm$5e-4} & \textbf{0.808} \scriptsize{$\pm$6e-3} & \textbf{0.0431} \scriptsize{$\pm$4e-4} & \textbf{0.967} \scriptsize{$\pm$3e-3} & \textbf{0.0361} \scriptsize{$\pm$3e-4} \\
    \hline
  \end{tabular}
  \label{tab:results_main}
\end{table*}

\subsubsection{Evaluation}
\label{sssec:sp_metrics}
We evaluated the performance of the preset encoders using two complementary metrics: the average $L^1$ error and the mean reciprocal rank (MRR). Each metric captures a different aspect of how well the preset encoder represents the synthesizer presets.

The average $L^1$ error, or mean average error (MAE), which should be minimized, measures the absolute difference between the preset representation produced by the preset encoder and the actual audio representation obtained by rendering the preset with the real synthesizer and processing it through the pretrained audio model. It focuses on fine-grained accuracy, where a lower $L^1$ error indicates that the proxy is closely aligning with the true audio features.

The MRR, which takes values in $\left[0,1\right]$ and should be maximized, evaluates how well the system can identify the correct preset representation from a set of candidates, given a target audio representation. Specifically, it measures how effectively the correct preset is ranked among other possible presets in terms of similarity to the target audio representation. A higher MRR indicates better performance, as it reflects the system’s ability to place the correct preset closer to the top of the ranking. 
Thus, while these metrics are correlated -- lower $L^1$ error typically leads to better MRR scores -- the MRR provides a more global evaluation by assessing the system’s ability to differentiate between presets in the entire representation space, as opposed to the $L^1$ error, focuses on the accuracy of individual embeddings. The detailed computation of the MRR for training and hyperparameter optimization is provided in Appendix \ref{app:mrr}.

To assess the performance of the encoders on synthetic presets, the following steps were conducted: (i) randomly selecting 4096 presets from the test set; (ii) calculating the average $L^1$ error and MRR using $Q$=4096, $K$=4096 (see Appendix \ref{app:mrr}); and (iii) repeating the first two steps for 100 runs with different random seeds to determine the mean and standard deviation of the metrics.

\subsubsection{Training}
\label{sssec:sp_training}
Each preset encoder was trained for 30 epochs using the Adam optimizer and a cosine decay learning rate scheduler. The batch size was set to 256 for TMF and 512 for other models. The initial learning rate was set to the optimal value determined through hyperparameter optimization and gradually decreased to a minimum of 2e-6, with a scheduler restart at the 10th epoch.

Two model checkpoints were selected for each preset encoder: one based on MRR (with $Q$=256, $K$=512, as described in Appendix \ref{app:mrr}) and another based on the average $L^1$ error calculated over the validation set. For each model, we present the results of the checkpoint that performed best on the test sets of hand-crafted presets, as the differences were marginal and did not affect the overall trend or interpretation of the results.

\subsection{Results}
\label{ssec:sp_results}
\begin{table*}[t]
  \caption{Comparison between models on hand-crafted presets.
    The mean and standard deviation of the MRR and average $L^1$ error over 100 runs are reported, along with the percentage difference from the values obtained on synthetic presets. The results for TAL-NM were omitted due to the limited number of available hand-crafted presets. The best-performing models for both metrics are highlighted in bold.}
  \centering
  \begin{tabular}{@{\extracolsep{4pt}}lcccccccc@{}}
    \hline
    \multicolumn{1}{c}{} & \multicolumn{4}{c}{Dexed}             & \multicolumn{4}{c}{Diva}                                                                                                                                                          \\
    \cline{2-5} \cline{6-9}
    Model                & MRR ($\uparrow$)                      & Diff.                    & $L^1$ error ($\downarrow$)             & Diff.   & MRR ($\uparrow$)                       & Diff.   & $L^1$ error ($\downarrow$)             & Diff.   \\
    \hline
    MLP-OH               & 0.020 \scriptsize{$\pm$1e-3}          & -91.3$\%$                  & 0.1495 \scriptsize{$\pm$8e-4}          & +56.9$\%$ & 0.065           \scriptsize{$\pm$2e-3} & -69.9$\%$ & 0.1277 \scriptsize{$\pm$7e-4}          & +74.0$\%$ \\
    HN-OH                & 0.045 \scriptsize{$\pm$2e-3}          & -93.2$\%$                  & 0.1044 \scriptsize{$\pm$6e-4}          & +72.6$\%$ & 0.234           \scriptsize{$\pm$4e-3} & -61.3$\%$ & 0.0898 \scriptsize{$\pm$6e-4}          & +64.2$\%$ \\
    HN-PT                & 0.031 \scriptsize{$\pm$2e-3}          & -94.7$\%$                  & 0.1092 \scriptsize{$\pm$6e-4}          & +64.2$\%$ & 0.212           \scriptsize{$\pm$3e-3} & -62.2$\%$ & 0.0912 \scriptsize{$\pm$5e-4}          & +41.4$\%$ \\
    HN-PTGRU             & 0.085 \scriptsize{$\pm$3e-3}          & -89.3$\%$                  & 0.0912 \scriptsize{$\pm$6e-4}          & +79.9$\%$ & 0.317           \scriptsize{$\pm$4e-3} & -49.5$\%$ & 0.0782 \scriptsize{$\pm$5e-4}          & +47.3$\%$ \\
    TFM                  & \textbf{0.124} \scriptsize{$\pm$4e-3} & -85.8$\%$                  & \textbf{0.0798} \scriptsize{$\pm$4e-4} & +83.4$\%$ & \textbf{0.455} \scriptsize{$\pm$5e-3}  & -43.7$\%$ & \textbf{0.0665} \scriptsize{$\pm$4e-4} & +54.3$\%$ \\
    \hline
  \end{tabular}
  \label{tab:results_gen}
\end{table*}

The primary results are summarized in Table \ref{tab:results_main}. TFM consistently outperformed other models across all synthesizers, followed by HN-PTGRU, HN-OH, HN-PT, and MLP-OH.
The results indicate a strong correlation between the complexity of the model architecture and its performance. This suggests that more advanced architectures are better suited for approximating the audio embedding space. The negative correlation between MRR and $L^1$ error is an expected outcome, indicating that improved approximation of the audio embedding space enhances the ability to identify correct presets.

All models achieve much higher MRRs and lower $L^1$ errors on TAL-NM compared to Dexed and Diva, which was expected due to TAL-NM's reduced number of parameters and fewer interdependencies between them. 
Therefore, the subsequent analysis will primarily focus on Dexed and Diva, which account for most of the variance in the results.

The baseline MLP-OH exhibits the lowest performance across all synthesizers, with a difference of up to +274$\%$ in MRR compared to TFM on Dexed.

The superior performance of the HN-OH model over the baseline model, with increases of +190$\%$ and +180$\%$ in MRR on Dexed and Diva, respectively, can be attributed to its greater architectural depth, as both models employ the same input encoding method. This is consistent with the understanding that deeper networks can learn more complex representations and thus perform better on tasks requiring high-level feature extraction.

Despite employing a more sophisticated encoding strategy, HN-PT underperforms compared to HN-OH across all synthesizers. This suggests that expressiveness alone does not guarantee the effectiveness of an encoding strategy and that more complex architectures may be necessary to fully leverage the information captured by the embedding layer, as demonstrated by the improved performance of HN-PTGRU and TFM.

Two potential factors may contribute to HN-PT's poor performance. First, the high ratio of input dimension to hidden dimension (11072:512 for Diva, 6400:512 for Dexed) likely introduces a representational bottleneck, which could hinder the model's ability to efficiently process and utilize the information encoded in the input embeddings.
Second, reduced hidden and token dimensions in HN-PT could limit its capacity to learn from the input embeddings. However, increasing these dimensions leads to scalability issues caused by the flattening operation, which drastically increases the number of parameters in the input layer.

Incorporating a BiGRU layer in HN-PTGRU leads to improved performance, yielding +36$\%$ and +12$\%$ in MRR on Dexed and Diva, respectively.
This enhancement is likely due to the ability of the BiGRU to extract a fixed-length context vector from the sequence of tokens, which mitigates the representational bottleneck caused by the flattening operation in HN-PT.
Furthermore, HN-PTGRU achieves greater parameter efficiency in the input layer, allowing the use of larger token and hidden dimensions, which increase the model capacity and result in better performance, while remaining in the same order of magnitude in terms of parameter count.

Interestingly, both HN-PT and HN-PTGRU exhibited fluctuating $L^1$ errors on the validation datasets and during evaluation, as indicated by the reported standard deviation. This indicates that their performance may vary with different input presets, necessitating further analysis to identify the causes of these instabilities.

The success of TFM can be attributed to its attention mechanism, since both HN-PT and HN-PTGRU use the same input encoding method but do not achieve this level of performance. This is particularly beneficial on Diva, for which capturing the interdependencies between parameters is crucial for accurate representation, yielding +44$\%$ and +29$\%$ in MRR compared to HN-PT and FT-GRU, respectively. These results are on par with SPINVAE \cite{vaillantSynthesizerPresetInterpolation2023} which also shows the superiority of the transformer-based model.

\section{Hand-Crafted Presets Evaluation}
\label{sec:hc_eval}
\subsection{Experimental Design}
\label{ssec:hc_design}
The evaluation on hand-crafted presets mirrors that of synthetic presets, excluding the use of TAL-NM as mentioned in Sec.~\ref{ssec:datasets} due to the limited number of hand-crafted presets. In fact, this could have made the results inaccurate, especially the MRR, since the smaller preset pool (404 vs. 4096) makes identifying the correct preset easier, even randomly. Moreover, this resulted in identical runs, which reduce the significance of the findings.

\subsection{Results}
\label{ssec:hc_results}
Table \ref{tab:results_gen} presents the evaluation results on hand-crafted presets, revealing a significant decrease in MRR and an increase in $L^1$ error across all synthesizers and models. This indicates that the models struggle to generalize from synthetic to hand-crafted presets.
TFM consistently outperforms the other models across both evaluation metrics and all synthesizers while still exhibiting a significant drop in performance when compared to the evaluation on synthetic presets. The relative performance trends between models align with the results from the evaluation on synthetic presets, providing further validation of the performance rankings established in the previous evaluation.

\begin{figure}[t]
  \centerline{\includegraphics[width=\linewidth]{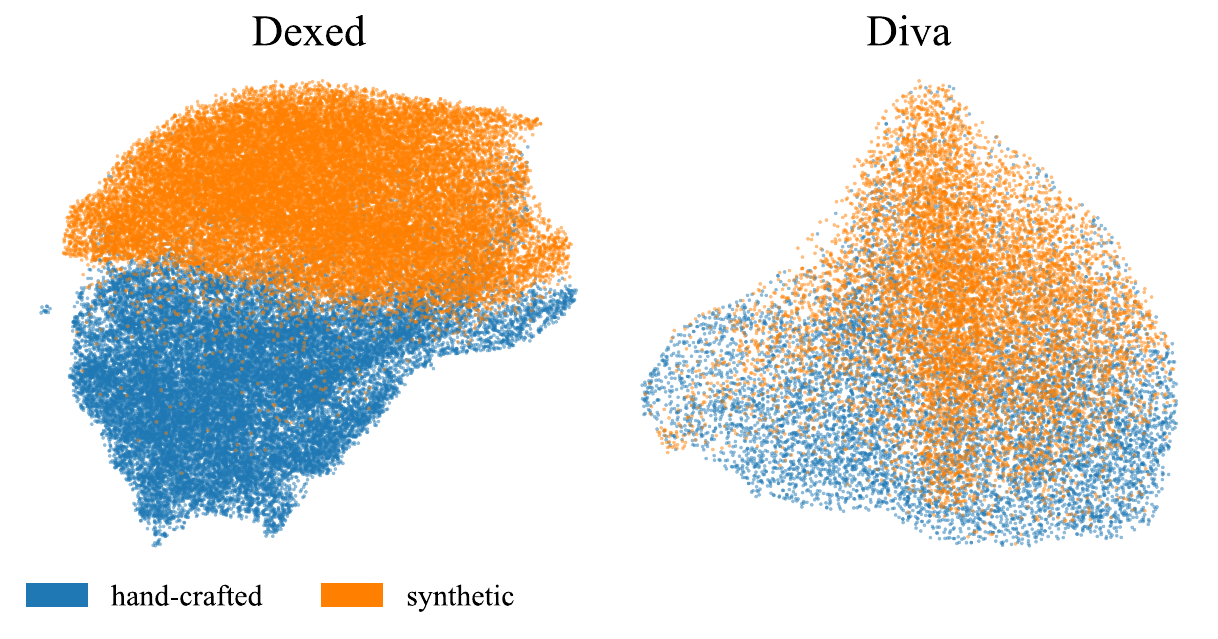}}
  \caption{\label{fig:umap_hc_vs_rnd}{UMAP projections of audio embeddings of both synthetic and hand-crafted presets. 30.0K randomly chosen synthetic and 27.6K hand-crafted presets were used for Dexed, while 10.0K randomly chosen synthetic and 9.0K hand-crafted presets were used for Diva.}}
\end{figure}

Interestingly, all models exhibit a higher performance drop on Dexed compared to Diva, suggesting that the random sampling process used to generate the training data is less favorable to Dexed than to Diva.
We hypothesize that the lack of generalization is due to a large subset of the synthesizer parameter space producing nonmusical or unpleasant sounds, which are mapped onto distinct regions of the audio embedding space.

This effect could be amplified by the increased complexity involved in programming a synthesizer to produce more intricate or musically appealing sounds, which is in part characterized by a greater interdependency between parameters and a narrower range of parameter values capable of producing such sounds. Thus, as the programming complexity of a synthesizer increases, the manifolds of presets corresponding to the more musically appealing sounds represent an increasingly smaller subset of the synthesizer parameter space.

This imbalance hinders the effective exploration of audio embedding space regions associated with more appealing sounds, such as hand-crafted presets. This occurs despite the large number of synthetic presets used during training, which offer a broader, yet less targeted, exploration of the synthesizer parameter space.
Consequently, hand-crafted and synthetic presets tend to be mapped onto increasingly distinct and non-overlapping subsets of the audio embedding space as the synthesizer becomes more complex.
Thus, while the preset encoder performs well across a broad subset of the parameter space, its efficacy diminishes with hand-crafted presets because it fails to extrapolate to regions of the audio embedding space unseen during training.

To support this hypothesis, the UMAP projections of the audio representations of both synthetic and hand-crafted presets for Dexed and Diva synthesizers were computed, as illustrated in Fig.~\ref{fig:umap_hc_vs_rnd}. Furthermore, we computed the Wasserstein distance between both clusters to estimate their proximity for each synthesizer. The Wasserstein distance is well-suited for this purpose as it quantifies the minimal 'cost' required to transform one distribution into another. The projections clearly show that synthetic and hand-crafted presets form distinct clusters on Dexed, while they are more intertwined on Diva. The Wasserstein distances were 3.52 for Dexed and 2.41 for Diva, which is consistent with the visual observations. This difference might be attributed to the high nonlinearity of FM synthesis used in Dexed, which demands more precise parameter tuning to achieve musical sounds, compared to the more linear and well-behaved synthesis techniques used in VA synthesizers \cite{jaffeTenCriteriaEvaluating1995}. This suggests that models trained on synthetic presets for Dexed are more prone to poor generalization than those trained on Diva, aligning with the results in Table \ref{tab:results_gen}.

These results underline the necessity of further research to mitigate the generalization gap between hand-crafted and synthetic presets. This includes refining the synthetic preset generation process to better model the joint distribution of synthesizer parameters \cite{figueiraSurveySyntheticData2022} and investigating domain adaptation techniques \cite{gongDLOWDomainFlow2019}.

\section{Sound Matching Downstream Task}
\label{sec:ssm_eval}

\subsection{Experimentation Design}
\label{ssec:ssm_design}
\begin{figure}[!ht]
\centering
\begin{subfigure}{0.5\textwidth}
    \includegraphics[width=\textwidth]{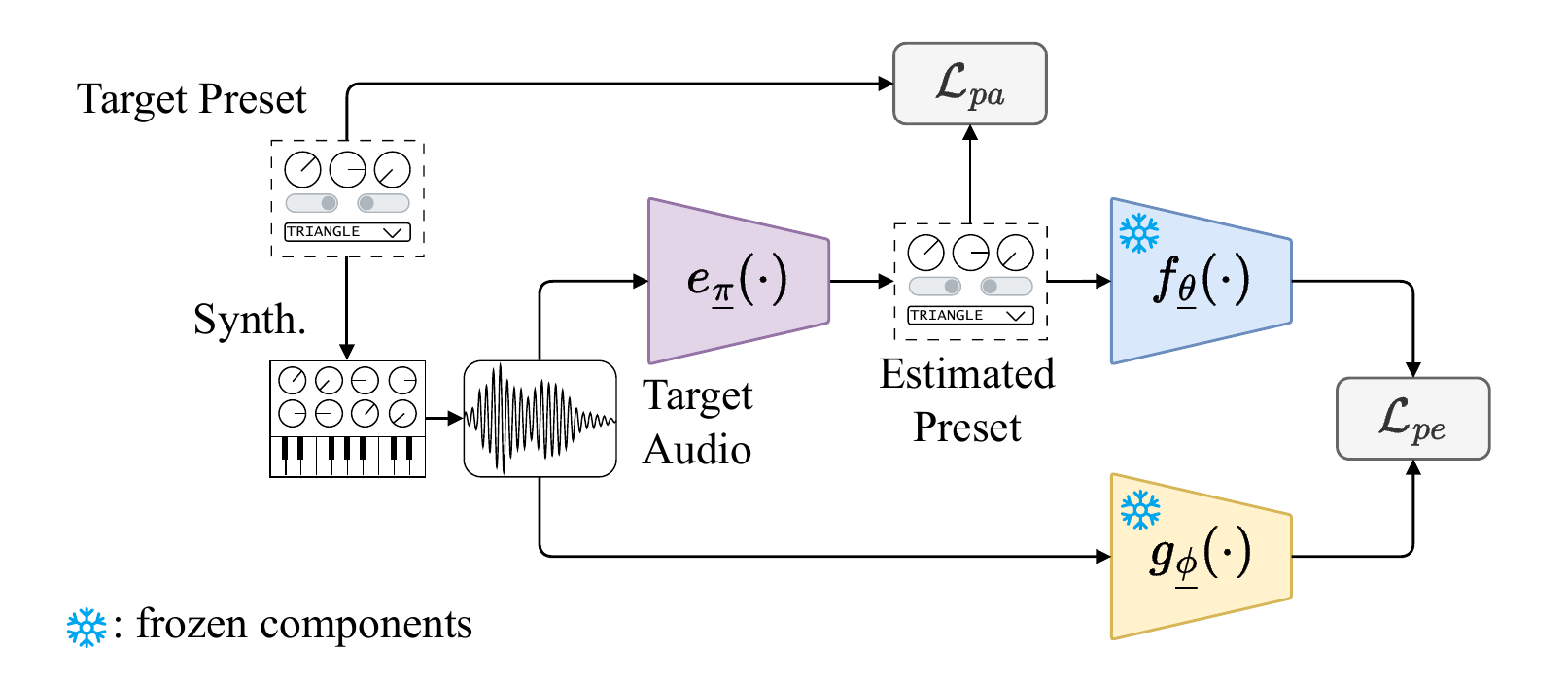}
    \caption{}
    \label{fig:ssm_train}
\end{subfigure}
\hfill

\begin{subfigure}{0.5\textwidth}
    \includegraphics[width=\textwidth]{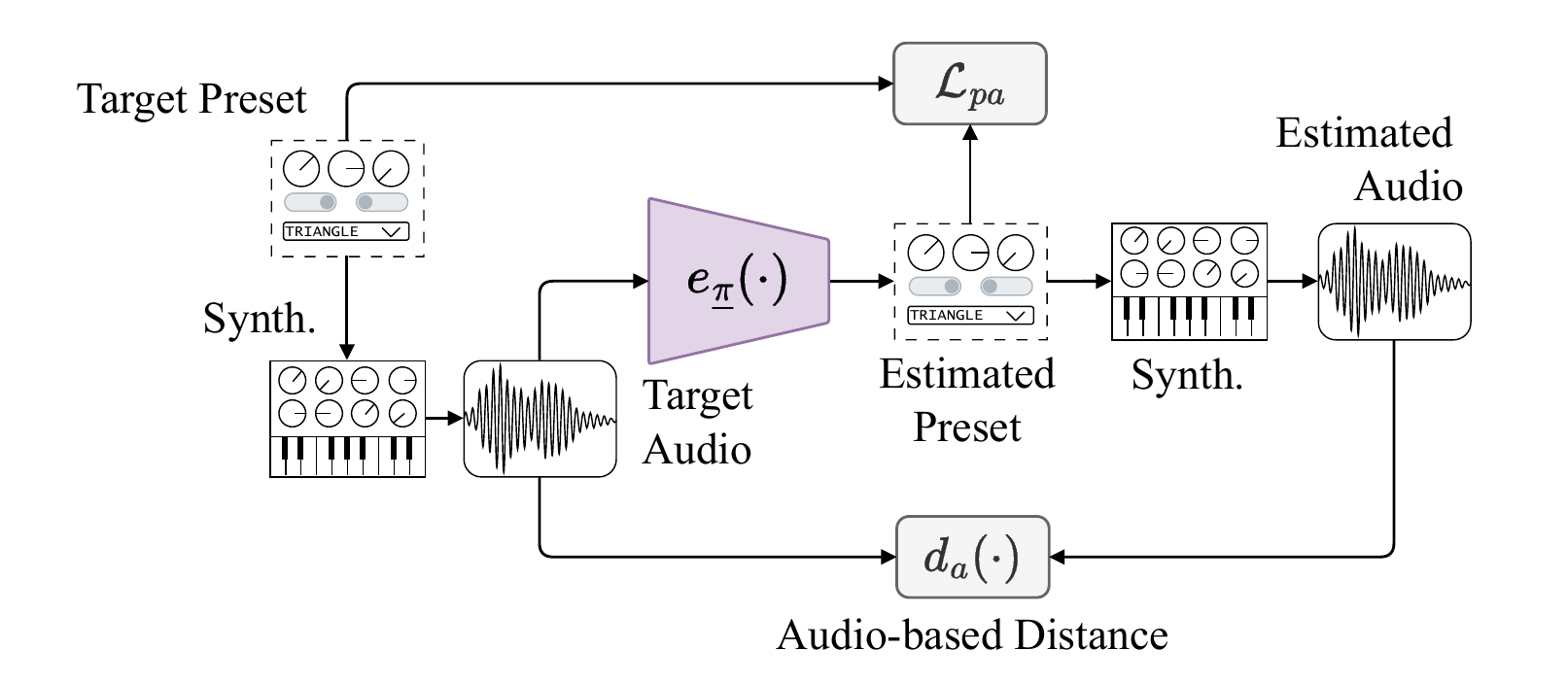}
    \caption{}
    \label{fig:ssm_test_id}
\end{subfigure}
\hfill
\begin{subfigure}{0.5\textwidth}
    \includegraphics[width=\textwidth]{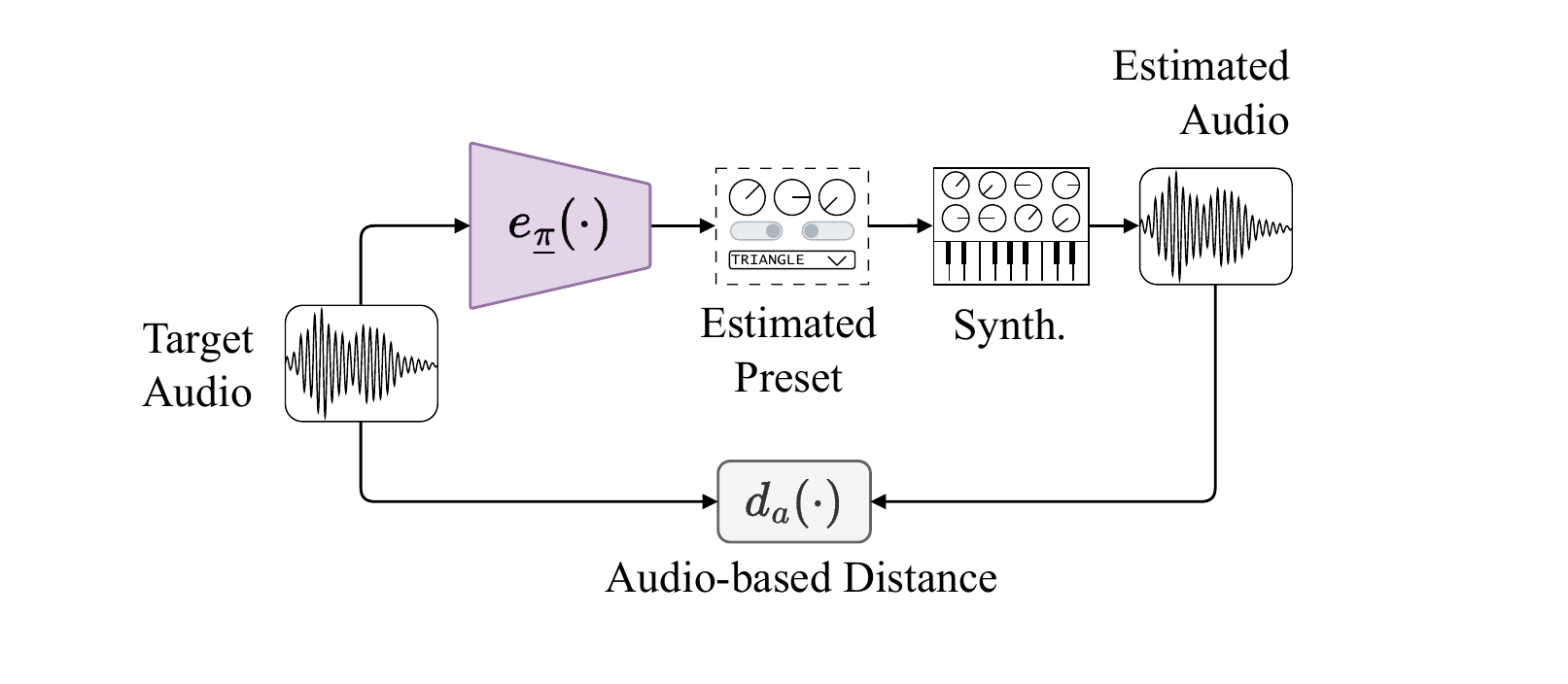}
    \caption{}
    \label{fig:ssm_test_ood}
\end{subfigure}
        
\caption{Overview of the training (a), in-domain validation and evaluation (b), and out-of-domain evaluation (c) pipelines for the SSM evaluation task. The estimator network is denoted as $e_{\underline{\pi}}$, while the preset encoder and pretrained audio model are denoted, as previously, by $f_{\underline{\theta}}$ and $g_{\underline{\phi}}$, respectively. $\mathcal{L}_{\text{p}}$ denotes the parameter loss function, while $\mathcal{L}_{\text{a}}$ denotes the audio embedding loss function. It is important to note that the preset encoder is only used during training and is replaced by the actual synthesizer during validation and evaluation stages.}
\label{fig:ssm_pipeline}
\end{figure}

To demonstrate the integration of a preset encoder into a nASP pipeline and assess its potential benefits, we evaluated its performance on a Synthesizer Sound Matching (SSM) downstream task using Dexed and Diva. The objective of the SSM task was to infer the set of synthesizer parameters that best approximates a given target audio. This section outlines our approach.

Regarding the preset encoder, we selected the TFM architecture, which yielded the best results, and pretrained it on a reduced dataset of 2.5M synthetic presets. This reduction was motivated by the diminishing returns observed using a larger dataset. This allowed us to replace the previous pretrained audio model -- \texttt{mn04} -- with the best-performing one -- \texttt{mn20} -- while maintaining a similar storage size, i.e., 7.68 GB for \texttt{mn04}@10M vs. 10.72 GB for \texttt{mn20}@2.5M. We then finetuned the preset encoder on a training set of hand-crafted presets, using the datasets presented in Sec.~\ref{sssec:hc_dataset} that were partitioned into an 80-10-10 train-validation-test split.

For the estimator network, which infers synthesizer parameters from input audio, we used the CNN-based architecture from \cite{barkanInversynthIISound2023}. The network takes as input 128-bin Mel spectrogram representations of the synthesized presets, computed from an FFT of size 1024 with a hop size of 512 and an input sample rate of 32 kHz. The estimator network was trained on in-domain sounds for 600 epochs using a loss function consisting of two terms. The first term is a parameter loss, consisting of $L^1$ loss for numerical parameters and cross-entropy loss for binary and categorical parameters, and an audio embedding loss, computed as the $L^1$ distance between the representation of the estimated preset -- produced by the preset encoder -- and the target audio's representation from the pretrained audio model. The training pipeline is illustrated in Fig. \ref{fig:ssm_train}.

\begin{table*}[t]
  \caption{Comparison of loss scheduling strategies on the test set of in-domain sounds. The best-performing models for each metric are highlighted in bold.}
  \centering
  \begin{tabular}{@{\extracolsep{0pt}}llccccccc@{}}
    \hline
  & & Num. $L^1$ ($\downarrow$) & Bin. Acc. ($\uparrow$) & Cat. Acc. ($\uparrow$) & STFT ($\downarrow$) & mSTFT ($\downarrow$) & Mel ($\downarrow$) & MFCCD ($\downarrow$) \\
\hline
\multirow{3}{*}{Dexed} & PLoss      & \textbf{0.074} & -              & \textbf{0.631} & 1.598          & 1.538          & 1.580          & 2.264 \\
                       & Mix        & 0.082          & -              & 0.617          & 1.487          & 1.428          & \textbf{1.454} & 2.085 \\
                       & Switch     & 0.082          & -              & 0.621          & \textbf{1.484} & \textbf{1.427} & 1.458          & \textbf{2.076} \\
\hline
\multirow{3}{*}{Diva}  & PLoss      & \textbf{0.117} & \textbf{0.848} & 0.364          & 2.058          & 1.984          & 2.206          & 2.318 \\
                       & Mix        & 0.129          & 0.845          & \textbf{0.367} & \textbf{1.779} & \textbf{1.722} & 1.825          & 1.955 \\
                       & Switch     & 0.177          & 0.765          & 0.331          & 1.811          & 1.756          & \textbf{1.815} & \textbf{1.892} \\
\hline
\end{tabular}
\label{tab:ssm_id}
\end{table*}

\begin{table}[t]
  \caption{Comparison of loss scheduling strategies on the test set of out-of-domain sounds. Since these sounds were not produced by the synthesizers at hand, parameter-based metrics were not computed. The best-performing models for each metric are highlighted in bold.}
  \centering
  \begin{tabular}{@{\extracolsep{0pt}}llcccc@{}}
    \hline
  & & STFT & mSTFT & Mel & MFCCD \\
\hline
\multirow{3}{*}{Dexed}  & PLoss     & 2.520          & 2.446          & 2.859          & 4.126 \\
                        & Mix       & \textbf{2.278} & \textbf{2.196} & \textbf{2.662} & \textbf{3.867} \\
                        & Switch    & 2.309          & 2.226          & 2.692          & 3.911 \\
\hline
\multirow{3}{*}{Diva}   & PLoss     & 2.197          & 2.132          & 2.856          & 3.763 \\
                        & Mix       & 1.912          & 1.850          & 2.392          & 3.347 \\
                        & Switch    & \textbf{1.848} & \textbf{1.773} & \textbf{2.168} & \textbf{3.224} \\
\hline
\end{tabular}
\label{tab:ssm_ood}
\end{table}

We evaluated three different loss scheduling configurations to assess the benefit of adding an audio embedding loss, inspired by \cite{masudaImprovingSemiSupervisedDifferentiable2023}: \begin{itemize}
    \item \texttt{PLoss}. Only the parameter loss is used and serves as a baseline.
    \item \texttt{Mix}. In this configuration, the parameter loss is applied for the first 200 epochs, then the audio embedding loss is gradually introduced over the next 200 epochs, and the estimator network is trained for the remaining 200 epochs using both parameter and audio embedding losses.
    \item \texttt{Switch}. This approach is similar to the previous one, but fully transitions from parameter loss to audio embedding loss, resulting in the estimator network being trained exclusively with audio embedding loss during the final 200 epochs.
\end{itemize}

As auditory evaluation metrics, we utilized a spectrogram and a 128-bin Mel spectrogram, both computed with an FFT size of 1024 and a hop size of 256, as well as a multiresolution spectrogram. All metrics are based on the $L^{1}$ distance, incorporating log-scaling and spectral convergence terms, and computed using the auraloss library \cite{steinmetz2020auraloss}. Additionally, we employed a 40-band Mel-Frequency Cepstral Coefficients Distance (MFCCD), computed from a 128-bin Mel spectrogram and measured using the $L^{1}$ distance.
For evaluation metrics based on parameter values, we used the $L^1$ distance for numerical parameters and the accuracy for categorical and binary parameters. 
For validation and evaluation, we replaced the preset encoder with the actual synthesizer, as shown in Fig. \ref{fig:ssm_test_id} for in-domain sounds and Fig. \ref{fig:ssm_test_ood} for out-of-domain sounds.
For each model, we selected the checkpoint that minimized the average of the audio-based metrics on the validation set, scaling the MFCCD to match the range of other metrics.

We acknowledge that our choice of evaluation metrics may appear contradictory to the results in Sec.~\ref{am_results}. However, we opted for hand-crafted representations to avoid overly optimistic performance estimates that could arise from using the same distance metric employed in the audio embedding loss. Additionally, selecting an alternative learned representation seemed arbitrary, and this choice ensures comparability with prior work that employs similar features, such as \cite{barkanInversynthIISound2023, masudaImprovingSemiSupervisedDifferentiable2023}.

The six models -- three loss schedules × two synthesizers -- were evaluated on both a test set of in-domain sounds and an additional test set of out-of-domain sounds containing the 12,678 examples from the validation set of the NSynth dataset \cite{engelNeuralAudioSynthesis2017a} which were upsampled to 32 kHz to match the sample rate used during the training of the estimator network. This allowed us to assess the benefits of incorporating a neural proxy into the nASP training process.

\subsection{Results}
\label{ssm_results}

The results of the SSM evaluation task for both in-domain and out-of-domain sounds are shown in Tables \ref{tab:ssm_id} and \ref{tab:ssm_ood}, respectively. Audio examples are available in the accompanying source code repository.

Overall, models trained using the \texttt{Switch} and \texttt{Mix} strategies outperform those trained using the \texttt{PLoss} strategy in auditory metrics for both in-domain and out-of-domain sounds. This suggests that incorporating audio embedding loss -- and consequently the preset encoder -- into the SSM training pipeline improves performance compared to relying solely on parameter loss. 
Although these models approximate the original sound through alternative synthesizer configurations, they generate presets that more closely resemble the target sound compared to those produced using the \texttt{PLoss} strategy.
These results align with previous findings \cite{barkanInversynthIISound2023,masudaImprovingSemiSupervisedDifferentiable2023,uzradDiffMoogDifferentiableModular2024}, where the inclusion of an audio-based loss function alongside a parameter loss led to improved performance. In contrast to the aforementioned studies, which use large, hand-crafted representations, our results suggest that improved performance can still be achieved using a relatively compact representation that integrates both hand-crafted and learned features -- a promising direction for efficient representations in nASP systems.
Moreover, the models demonstrate better generalization on out-of-domain sounds for the Diva synthesizer, despite the smaller dataset. This may be due to the more intuitive nature of VA synthesis compared to FM synthesis or the similarity of the dataset to NSynth. 

The performance difference between the \texttt{Mix} and \texttt{Switch} configurations is unclear and seems to depend on the specific synthesizer used. Future work should explore the impact of incorporating out-of-domain sounds into the training set, determine the optimal balance between perceptual and parameter losses, and investigate alternative schedules and training strategies.

Finally, although listening to the predicted presets supports the relative trends observed in the quantitative evaluations, achieving high imitation quality remains difficult, underscoring that the sound matching task is still challenging.

\section{Limitations}
\label{sec:limitations}
While learned representations show promise, there are several limitations to consider, particularly regarding practicality and task-specific suitability.

First, Sec.~\ref{am_results} demonstrated that the most effective and compact representations combined both hand-crafted and learned features. Removing the hand-crafted features from the \texttt{mn04} model in Sec.~\ref{sec:sp_eval} led to a significant performance drop. Additionally, the best-performing representations relying solely on learned features -- such as those from OpenL3 and AudioMAE -- tended to have significantly larger embedding sizes and lower throughput, which raises concerns about their practicality in applications with constrained resources.
In fact, unlike hand-crafted representations, which can be computed on the fly without the need for large amounts of stored data, learned representations often involve more complex algorithms, lower throughput, and require storing weights. In such cases, pre-computation and storage of the embeddings become necessary, which could be problematic in applications with limited storage or computational resources. Potential directions to tackle this problem would be to better explore the benefit of hybrid representations and methods for model optimization \cite{liuSurveyModelCompression2025}.

Moreover, not all learned representations are equally suited to the target task. For instance, models designed for neural audio compression performed poorly in our evaluation (see M2L and DAC in Sec.~\ref{am_results}), likely due to a mismatch between their training objectives and the perceptual ranking task. This underscores the importance of evaluating audio models with respect to the target task.

\section{Conclusion}
\label{sec:conclusion}
In this study, we proposed a method for training neural proxies for black-box synthesizers by mapping presets to an audio embedding space defined by a pretrained audio model. We demonstrated that transformer-based models outperform other encoder architectures on both synthetic and hand-crafted presets across three synthesizers.
Furthermore, we compared representations derived from various pretrained audio models in the context of nASP systems, highlighting the potential benefit of including both hand-crafted and learned features to obtain compact yet effective hybrid representations. We also discussed the limitations of relying exclusively on learned representations.
Additionally, we illustrated the integration of neural proxies into nASP systems through a synthesizer sound-matching downstream task, utilizing a compact, hybrid representation produced by the preset encoder as an audio embedding loss. The results were promising and consistent with findings from prior studies.
However, the generalization gap between synthetic and hand-crafted presets, as well as the integration of synthesizers into deep learning pipelines, remains a significant challenge. 

To improve the performance of nASP systems, future work should focus on developing more challenging benchmarks to assess the performance of learned and hand-crafted representations in nASP-related tasks. Additionally, research should explore the benefits of hybrid representations, as well as optimization techniques for developing compact and efficient representations suitable for low-resource settings. Finally, synthetic preset generation strategies should be investigated to address the data scarcity associated with hand-crafted presets.
Further investigations into the application of synthesizer proxies in nASP methods should focus on enabling broader control parameter coverage, including velocity and pitch, and on exploring alternative training schedules and strategies.

\section{Acknowledgments}
The authors wish to thank the anonymous reviewers for their feedback, which helped improve and clarify this manuscript; Stéphane Thunus for his thorough proofreading and feedback; Joseph Turian for his initial guidance on the topic and methodology; Gwendal le Vaillant for the useful discussion on SPINVAE; the team of TU Berlin's HPC Cluster for providing the computational facilities; and Ana-Marija Cvitic for her support.

\balance
\bibliographystyle{abbrvnat}
\bibliography{arxiv.bib}

\break

\appendix
\nobalance
\renewcommand{\thesubsection}{A.\arabic{subsection}}
\section*{Appendix}

\subsection{Overview of Pretrained Audio Models for Ranking Evaluation}
\label{app:am_descr}
This section provides an overview of the pretrained audio models evaluated in Sec.~\ref{sec:am_select}. For more detailed information, the reader is referred to the original papers.

\emph{Audio MAE} \cite{NEURIPS2022_b89d5e20}. A Vision Transformer (ViT)-based masked autoencoder trained on AudioSet, denoted as \texttt{AudioMAE}. The final representation for each audio frame is computed by averaging the outputs of the last four transformer blocks. The input sample rate is 16 kHz.

\emph{CLAP} \cite{wuLargescaleContrastiveLanguageAudio2024}. A multimodal audio-text model that utilizes the HTSAT transformer-based architecture as its audio backbone. The final representation corresponds to the projection of the time-averaged output of the audio submodel onto the shared audio-text embedding space. We evaluate two variants: one trained on music, speech, AudioSet, and LAION-Audio-630k, denoted as \texttt{CLAP-msa}, and another trained on music, speech, and LAION-Audio-630k, denoted as \texttt{CLAP-ms}. Detailed information about the specific music and speech datasets used to train these models is not publicly available. The input sample rate is 48 kHz.

\emph{Descript Audio Codec} \cite{NEURIPS2023_58d0e78c}. A CNN-based autoencoder built upon the VQ-GAN framework, trained on a large dataset comprising speech, music, and environmental sounds, denoted as \texttt{DAC}. For each audio frame, the final representation is the concatenation of projected latents from each stage (i.e., the continuous representation of the input before quantization). The input sample rate is 44.1 kHz.

\emph{EfficientAT} \cite{schmidLowComplexityAudioEmbedding2023}. A family of models based on MobileNetV3, with network widths scaled by various factors (e.g., \texttt{mn20}, corresponding to a scaling factor of 2.0), and trained on AudioSet using knowledge distillation from a transformer ensemble. For each audio frame, the output representation is produced by concatenating features from intermediate layers with the time-averaged input Mel spectrogram. The input sample rate is 32 kHz.

\emph{OpenL3} \cite{cramerLookListenLearn2019}. A CNN-based audio-video multimodal model trained on AudioSet. We use checkpoints from \cite{dasTorchopenl32021} trained on 256-bin Mel Spectrogram of the music subset of AudioSet to extract 6144-dimensional and 512-dimensional representations, denoted as \texttt{OpenL3-6144} and \texttt{OpenL3-512}, respectively. For each time frame, the final representation is the output of the audio submodel. The input sample rate is 48 kHz.

\emph{Music2Latent} \cite{pasiniMusic2LatentConsistencyAutoencoders2024}. A CNN-based diffusion autoencoder trained on the MTG Jamendo dataset and clean speech segments from DNS Challenge 4, denoted as \texttt{M2L}. For each audio frame, the final representation is obtained before the encoder bottleneck, with an input sample rate of 44.1 kHz.

\emph{PaSST} \cite{koutiniEfficientTrainingAudio2022}. A ViT-based model trained on AudioSet using Patchout. For each audio frame, the final representation is the concatenation of outputs for two different window lengths and the input’s flattened Mel spectrogram from either the large or small model, denoted as \texttt{PaSST-S} and as \texttt{PaSST-L}, respectively. The input sample rate is 32 kHz.

\emph{Baselines}. Three baselines were used to compare the performance of the pretrained audio models: (i) \texttt{mel128}, a 128-bin Mel spectrogram computed from an audio input with a sample rate 44.1 kHz using an STFT with a window size of 2048 and a hop size of 1024; (ii) \texttt{mfcc40}, Mel-frequency Cepstral Coefficients (MFCCs) of 40 bands computed from a 256-bin Mel spectrogram using the aforementioned STFT configuration; and (iii) \texttt{mstft}, a multi-resolution log-spectrogram computed from window sizes in $\{4096, 2048, 1024, 512, 256, 128, 64\}$ and hop sizes in $\{2048, 1024, 512, 256, 128, 64, 32\}$.

\subsection{Mean Reciprocal Rank Computation for the Evaluation on Synthetic and Hand-Crafted Presets}
\label{app:mrr}
The computation of the MRR, introduced in Sec.~\ref{sssec:sp_metrics} and used as a metric for the evaluation and hyperparameter optimization of the preset encoders, is described as follows.

Given a dataset of presets $\mathcal{D}:=\{\underline{x}^{(i)}\}_{i=1}^p$, we first randomly sample $K \leq p$ presets $\{\underline{x}^{(i)}\}_{i \in I}$ and a target index $q \in I$, where $I$ denotes the index set of the sampled presets such that $\vert I\vert=K$.
Then, we compute the vector of indices,
\begin{equation}
  \underline{\kappa}^{(q)}:=\operatorname{argsort}\left(\left[\left\|f_{\underline{\theta}}\left(\underline{x}^{(i)}\right)-g_{\underline{\phi}}\left(\underline{x}_a^{(q)}\right)\right\|_1\right]_{i\in I}\right) \in \mathbb{N}^K
\end{equation}
sorting the vector of $L^1$ distances between the preset representations $\{f_{\underline{\theta}}(\underline{x}^{(i)})\}_{i \in I}$ produced by the preset encoder $f_{\underline{\theta}}$ and the target audio representation $g_{\underline{\phi}}(\underline{x}_a^{(q)})$ produced by the pretrained audio model $g_{\underline{\phi}}$ in ascending order.

Next, we determine the rank $r^{(q)} \in \{j\}_{j=1}^{K}$ of the matching preset representation $f_{\underline{\theta}}(\underline{x}^{(q)})$ such that $I[\underline{\kappa}^{(q)}[r^{(q)}]]=q$.
We repeat this process $Q$ times, either by iterating over all possible $q \in I$ for model evaluation (see Sec.~\ref{sssec:sp_metrics}), or by sampling presets from $\mathcal{D}$ without replacement until all presets have been sampled during the validation stage of the training and hyperparameter optimization, detailed in Sec.~\ref{sssec:sp_training} and Appendix \ref{app:sp_hpo}, respectively. The MRR is finally computed as
\begin{equation}
  \operatorname{MRR}:=\frac{1}{Q} \sum_{q \in I_q} \frac{1}{r^{(q)}} \in[0,1]
\end{equation}
where $I_q$ is the index set of all target presets such that $\vert I_q \vert = Q$. The ideal outcome is for the preset representation $f_{\underline{\theta}}(\underline{x}^{(q)})$ to be the closest to $g_{\underline{\phi}}(\underline{x}_a^{(q)})$ among all preset representations, ideally achieving $r^{(q)}=1$.

\subsection{Hyperparameter Optimization}
\label{app:sp_hpo}
The hyperparameter optimization (HPO) for the backbone architectures and their corresponding optimizers was performed using the Optuna framework \cite{akibaOptunaNextgenerationHyperparameter2019}. This process involved two stages: an initial exploration phase with a Quasi-Monte Carlo (QMC) sampler, followed by an exploitation phase using a Tree-structured Parzen Estimator (TPE) sampler. Each HPO trial consisted of sampling a set of hyperparameters that was employed to train a model for a single epoch, with batch sizes of 256 for TMF and 512 for other models. The objective was to maximize the MRR on the validation set by choosing $Q$=256, $K$=256 (see Appendix \ref{app:mrr}).

Architecture and optimizer hyperparameters were optimized on TAL-NM and Diva over 128 QMC trials and 75 TPE trials. Architecture hyperparameters were selected to ensure that all models had a similar number of parameters. TFM's hyperparameters were tuned only on TAL-NM since the computational cost of HPO on Diva was prohibitively high. For Dexed, only the optimizer parameters were optimized for all encoders over 64 QMC trials and 36 TPE trials.

\end{document}